\documentclass{PoS}

\title{Particle detection technology for space-borne astroparticle experiments}

\ShortTitle{Particle detection technology for space-borne astroparticle experiments}

\author{\speaker{Martin Pohl} \\ 
        DPNC and CAP Genève\\
        Université de Genève\\
        Switzerland\\
        E-mail: \email{martin.pohl@unige.ch}}

\abstract{I review the transfer of technology from accelerator-based equipment to space-borne astroparticle detectors. Requirements for detection, identification and measurement of ions, electrons and photons in space are recalled. The additional requirements and restrictions imposed by the launch process in manned and unmanned space flight, as well as by the hostile environment in orbit, are analyzed. Technology readiness criteria and risk mitigation strategies are reviewed. Recent examples are given of missions and instruments in orbit, under construction or in the planning phase. }

\FullConference{Technology and Instrumentation in Particle Physics 2014,\\
		2-6 June, 2014\\
		Amsterdam, the Netherlands}

\begin{document}

\section{Introduction}
The task of astroparticle detectors in space is to measure cosmic rays and photons. For cosmic rays, the composition and spectra are to be measured, typically in the energy range from 100 MeV to multi-TeV. The particle energy, mass and charge need to be determined
as well as the arrival time and direction, although the latter is less important since charged particles are diffused by cosmic magnetic fields. In this energy range, most of the cosmic rays are supposed to be of galactic origin~\cite{ref_Blasi}. For photons, typically in the X-ray to TeV energy range, the incoming direction allows to identify the astrophysical source. Energy, timing, as well as polarization degree and direction are obvious observables. Extended and point sources as well as diffuse components count among the many targets of observation~\cite{ref_Thierry}. In addition, like all particle physics experiments, astroparticle detectors need to be sensitive to the unknown, be it unusual components of radiation or unknown sources. The space environment is characterized by low rates and thus low occupancies of the detector, if energies below the region of interest are shielded. 

\begin{figure}
\hfill \includegraphics[height=0.49\textwidth]{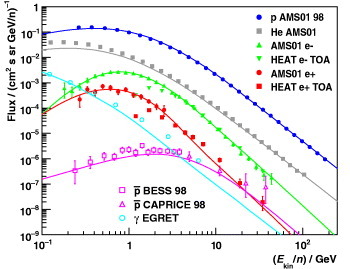} \hfill
\includegraphics[height=0.49\textwidth]{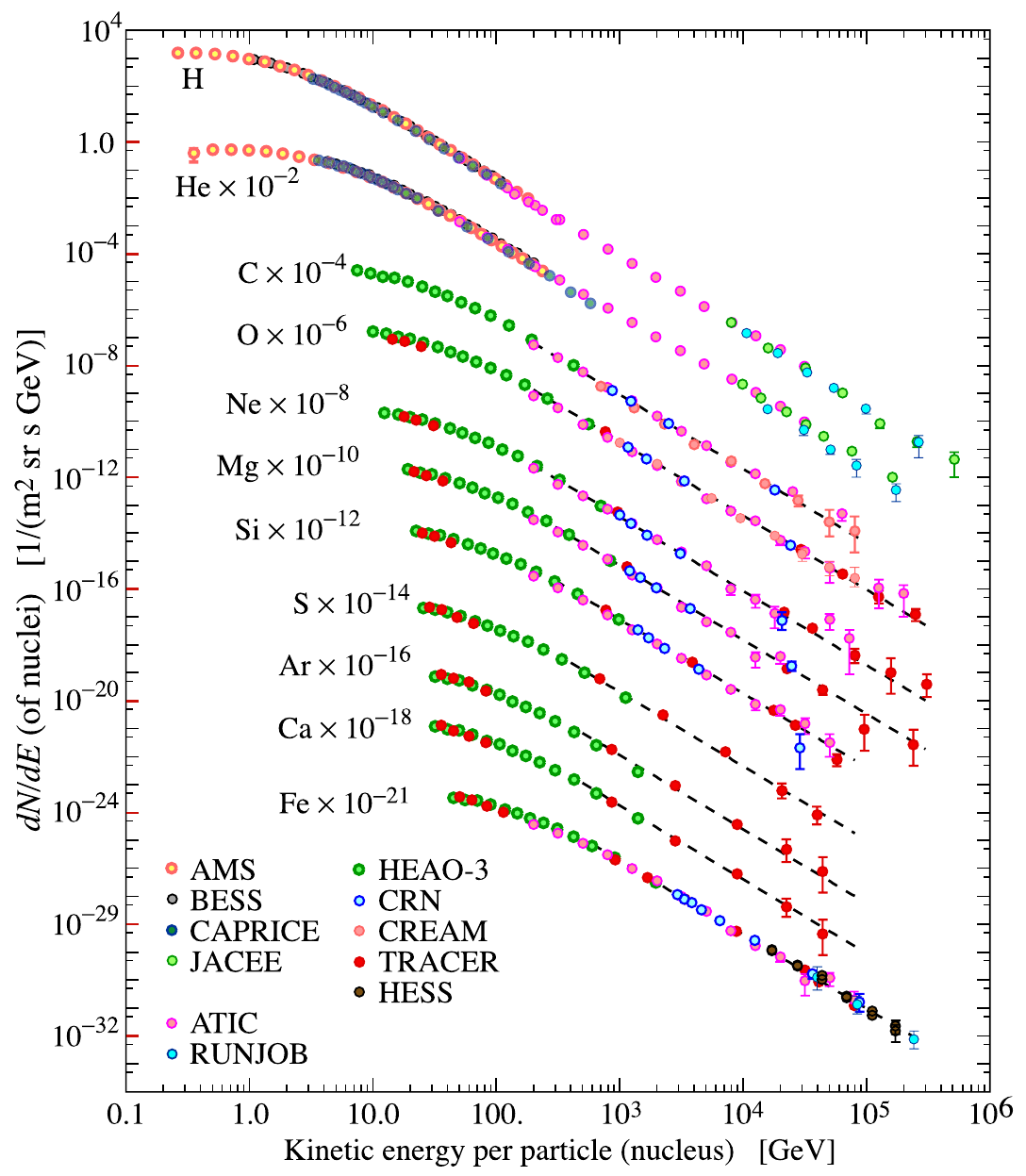} \hfill $ $ 
\caption{Left: Cosmic ray and photons fluxes in the 1 to 100  GeV range~\cite{ref_Beischer}. Right: Fluxes of nuclei from GeV to high energies~\cite{ref_PDG}.}
\label{fig_fluxes} 
\end{figure}

As shown in Fig.~\ref{fig_fluxes}, the fluxes of cosmic radiation fall like power laws, with a typical decrease in intensity by almost three orders of magnitude per decade in energy. It is composed of about 87\% protons, 9\% He nuclei, a few percent heavy nuclei and even fewer electrons, positrons, antiprotons and photons. It is thus obvious that size matters, the acceptance in terms of surface and field-of-view and the exposure time determine the energy reach of the detector. On the other hand, rocket launch and operation in the hostile space environment cause severe limitations in terms of size, weight and power consumption, and demanding requirements for environmental resistance and reliability. 

\section{Launch and In-Orbit Requirements}
The mechanical requirements on space hardware are dominated by the harsh launch process. High levels of acoustic noise, static and vibrational loads are encountered. Moreover, pyro-shocks during the separation of rocket components can reach levels of several thousand $g$ and high frequencies in unmanned space flight. It is thus mandatory to qualify equipment for both longitudinal and lateral vibrations and shocks, and to determine resonant frequencies. Qualification levels usually include a large margin of safety compared to actual launch conditions. Other requirements concern interactions with the space craft and other payloads, like cleanliness, particle and chemical contaminations and electromagnetic compatibility. Details of the qualification requirements are given in the documentation of each launch vehicle, for popular ones see~\cite{ref_manuals}.  

The in-orbit environment is unusually hostile for sensitive particle detectors. Micrometeorites and orbiting debris may impact at the speed and mass of a gun projectile. Atomic oxygen may corrode surfaces. Charging by plasma is an issue, especially when transiting the Earth radiation belts. Irradiation~\cite{ref_Bourdarie} by protons and electrons trapped in the Earth magnetosphere, by solar particles and cosmic rays themselves may cause latch-up or single event effects. Usage of modern electronic components or sensors thus often requires qualification for the expected environment. Dose rates can be predicted using the SPENVIS package~\cite{ref_SPENVIS}.

The thermal environment in space often presents the most demanding challenge, since it depends on many variables. Orbital properties determine the periodic day/night transition. The angle $\beta$ between the orbital plane and the solar vector determines illumination conditions. Radiators and solar panels of the space craft may shadow the payload. The payload attitude may suddenly change by space craft maneuvers. All of this introduces thermal changes with very different time scales, from minutes to months. The result may be thermo-mechanical deformations requiring dynamic alignment. Properties of electronic components like noise levels and gains may shift as a function of temperature, requiring frequent calibration. And ultimately, damaging effects may occur if excursions outside the operational or even survival temperature range of the equipment are encountered. Thermal control can in the best case be implemented passively by heat conduction towards radiators. Active thermal control can require pumps and pressurized systems which are more difficult to design for space applications~\cite{ref_Vanes}. Payload operations on ground may not always be able to control the thermal environment at any given time. It is thus wise to implement automatic safeguard procedures.

The most demanding requirements -- for someone used to develop hardware and software for ground-based particle and astroparticle physics -- may well be the extensive documentation mandatory throughout the design, qualification, construction and acceptance process. Exhaustive documentation is required on the design, manufacture and assembly, as well as the verification procedures and their results. All of this down to the smallest structurally relevant detail. A complete traceability of components, non-compliances and incidents is also part of the process. All-in-all one may expect several 10k pages of paperwork, even for a modest size mission.   

\section{Reliability and Risk Mitigation}
\begin{figure}
\begin{center}
\includegraphics[width=0.49\textwidth]{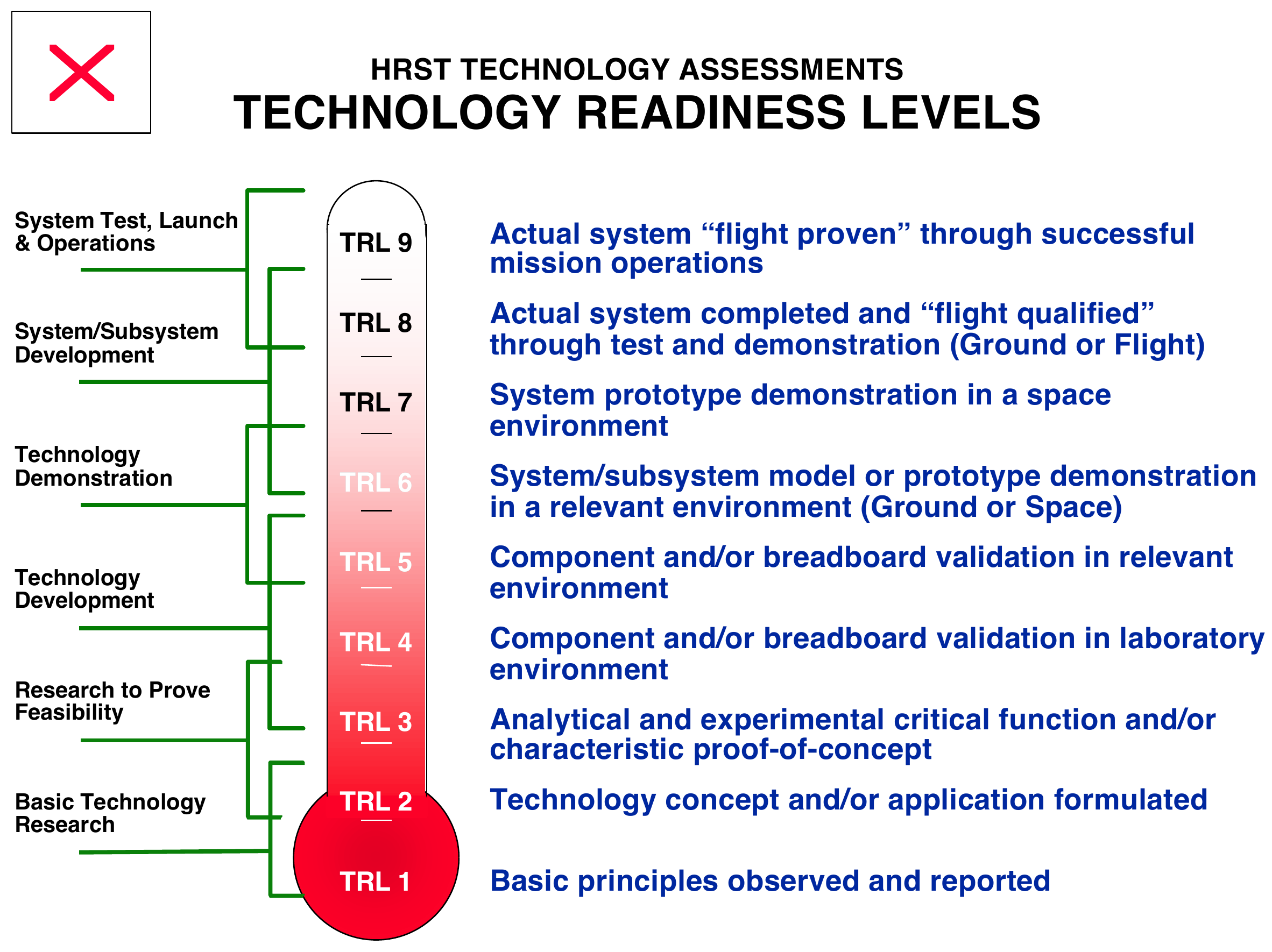}
\end{center}
\caption{Definition of Technology Readiness Levels and operations required for transiting them~\cite{ref_TRL}.}
\label{fig_TRL} 
\end{figure}

With few exceptions, equipment in space remains inaccessible during a mission which may last for up to a decade.  It is thus of utmost importance to assess the reliability of equipment under space conditions. Space agencies like NASA and ESA use the concept of Technology Readiness Level (TRL)~\cite{ref_TRLISO} to judge the maturity of components and subsystems in a risk analysis. Fig.~\ref{fig_TRL} summarizes the nine levels of TRL, which range from the establishment of basic principles (TRL 1) to flight proven equipment through successful missions (TRL 9). It is to be noted that successful laboratory tests of prototype equipment, usually acceptable to approve ground based particle physics experiments, only take you to TRL 4. Traversing the rest of the scale requires subsystem and system level qualification for space environment, which is a costly and tedious procedure. For this reason, you will often notice a considerable delay in porting new technologies from ground based equipment to space instruments. 

Risk mitigation is also required because the cost of the launch may be comparable to or exceed the cost of the instrument itself. Constructing a space particle detector may also be a once-in-your-lifetime experience because of the scarce flight opportunities devoted to science in general and to astroparticle physics in particular. It is therefore common to use established technology for detectors and electronics and to count on redundancy to increase reliability and lifetime. It is also important to enforce redundant measurements of crucial quantities by multiple detector components, creating synergy and reliability at the same time. Needless to say that exhaustive testing in all phases of design, development and construction is mandatory.

\section{Spectrometers and Calorimeters in Orbit}
\begin{figure}
\includegraphics[height=0.41\textwidth]{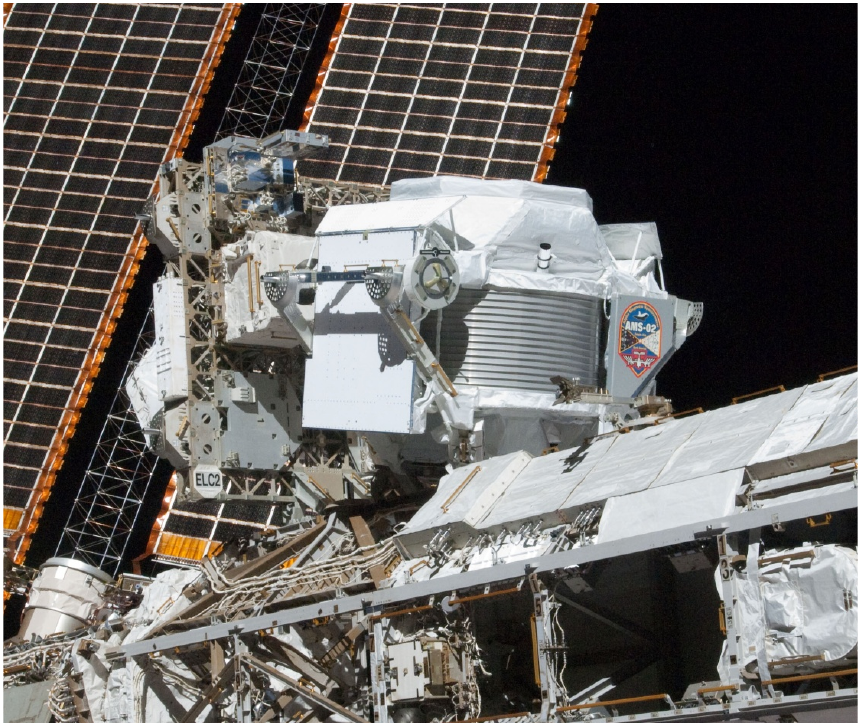} \hfill
\includegraphics[height=0.41\textwidth]{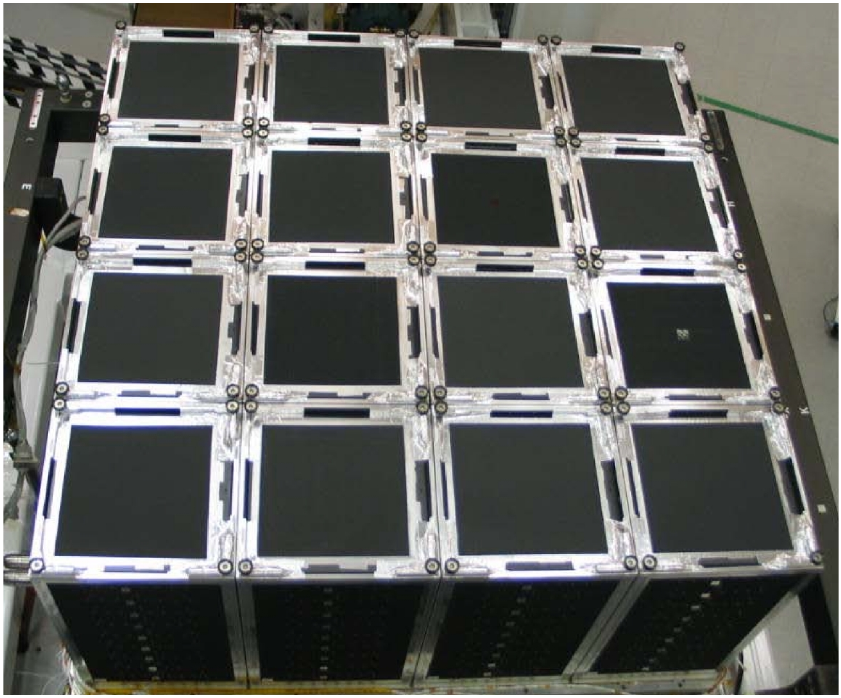}
\caption{Left: The AMS cosmic ray spectrometer on the ISS. Right: The Fermi Large Area Telescope with its 16 tower modules.}
\label{fig_AMS_Fermi} 
\end{figure}

Particle detector systems in space can be broadly categorized into spectrometers and calorimeters. Since size determines energy reach (and cost), it is logical that a smaller mission is often followed in time by a larger one. The spectrometers PAMELA~\cite{ref_PAMELA} and AMS~\cite{ref_AMS}, different in acceptance by two orders of magnitude, are examples of this sequence. For calorimeters, AGILE~\cite{ref_AGILE} and Fermi/LAT~\cite{ref_Fermi} follow a similar logic. Tab.~\ref{tab_AMS_Fermi} compare some rough figures of merit for the two larger instruments. As required by the low photon flux and accommodated more easily by calorimetry, the Fermi/LAT acceptance is larger than what a magnetic spectrometer like AMS can reach. Fig.~\ref{fig_AMS_Fermi} shows photographs of the two instruments.

\begin{table}
\begin{center}
\begin{tabular}{|l|c|c|}
\hline
								& AMS-02				& Fermi/LAT					\\ \hline
Acceptance $[ \mbox{m}^2\mbox{sr}]$	& 0.4 (tracker)			& 2.4 						\\
								& 0.1 (ECAL)			&							\\ \hline
Energy range 						& 100 MeV -- few TeV	& 20 MeV -- 300 GeV 			\\ \hline
Typical $\sigma_E/E$				& 2.5\%				& 9\% -- 18\%					\\ \hline
Typical $\sigma_\Omega$			& $0.14^\circ$ (tracker)	& $< 0.15^\circ$ for $E>10$ GeV 	\\
								& $< 1^\circ$ (ECAL)	&							\\ \hline
\end{tabular}
\end{center}
\caption{Rough figures of merit characterizing acceptance and performance for the leading cosmic ray spectrometer AMS and the leading general purpose tracking calorimeter Fermi/LAT.}
\label{tab_AMS_Fermi}
\end{table}

The tracker of the AMS spectrometer~\cite{ref_AMS_tracker} and the tracking sensors of the Fermi/LAT calorimeter~\cite{ref_Fermi_tracker}  inherit from the rich experience with silicon strip detectors at colliders. The conservative approach dictates the use of high resistivity single or double sided detectors, heavily used in LEP and LHC experiments. Likewise calorimeters rely on sampling technology inherited from collider detectors (Pb/scintillating fibers in AMS~\cite{ref_AMS_ECAL}, Si/W in Fermi/LAT~\cite{ref_Fermi}) and homogeneous detection materials like CsI~\cite{ref_Fermi,ref_AGILE_calo}, BGO~\cite{ref_DAMPE}, PbWO$_4$~\cite{ref_CALET} or Lyso~\cite{ref_HERD}. 

\begin{figure}
\begin{center}
\raisebox{8mm}{\includegraphics[width=0.58\textwidth]{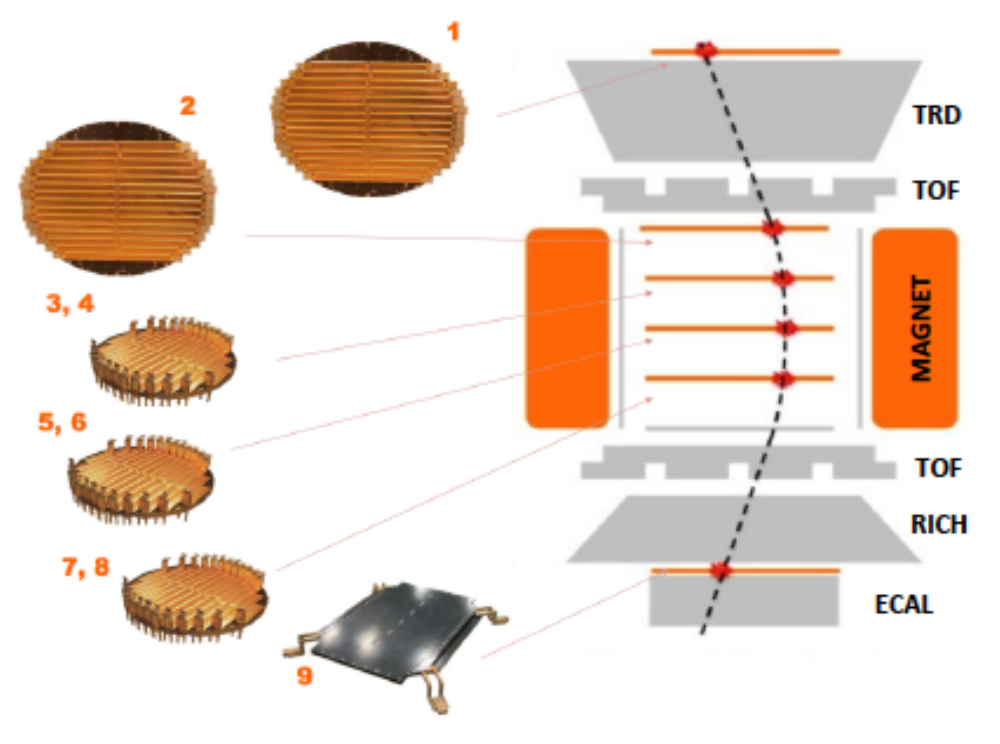}} \hfill
\includegraphics[width=0.40\textwidth]{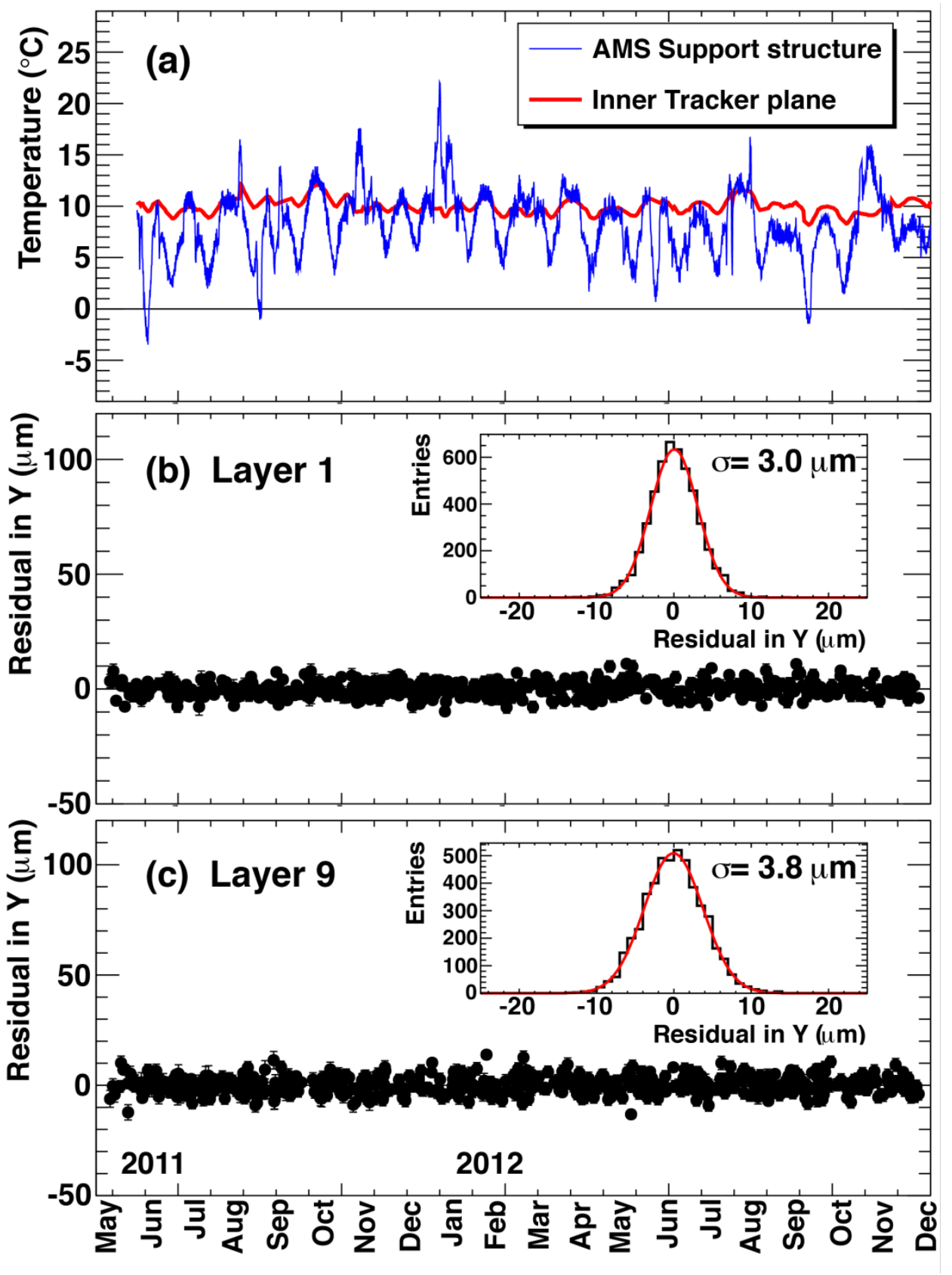}
\end{center}
\caption{Left: Sketch of the AMS set-up showing its subsystems and the location of the tracker planes.  
Right: Temperature sensor readings as a function of time (top) in the actively controlled region of AMS (red) and the passively controlled ones (blue). Result of the dynamic alignment procedure using cosmic ray tracks as a function of time for the topmost layer 1 (middle) and the bottom layer 9 (bottom). In both cases, stability to within a few micrometers is reached.}
\label{fig_AMS_shifts} 
\end{figure}

As pointed out before, redundancy and synergy are key to the long lifetime and systematic accuracy of astroparticle measurements in space. As an example, the spectrometer and electromagnetic calorimeter of AMS (see Fig.~\ref{fig_AMS_shifts}) measure momentum and energy of electrons and positrons redundantly, while a TRD~\cite{ref_AMS_TRD} distinguishes light from heavy particles. Likewise, the time-of-flight scintillator planes~\cite{ref_AMS_ToF}, RICH~\cite{ref_AMS_RICH} and tracker~\cite{ref_Pierre} all measure the particle charge simultaneously. 

An example of the varying thermal environment and its consequences is exhibited by the 
AMS spectrometer as shown in Fig.~\ref{fig_AMS_shifts}. While the inner volume of the magnet is actively cooled~\cite{ref_Vanes} and thus stable to the micron level, the outermost top and bottom planes regularly shift by as much as several hundred micrometers throughout thermo-mechanical deformation of the support structure. These shifts are tracked by dynamical realignment, using the about 10k cosmic ray tracks that AMS sees every minute. The stability of the resulting alignment, also shown in Fig.~\ref{fig_AMS_shifts}, 
allows to fully recover the required spatial resolution of a few microns at all times. 

\section{Imminent and Future Missions}
The next round of space-borne astroparticle detectors follows the logic of "bigger and better" but also incorporates novel ideas. In particular, high energy missions become more general purpose and combine extended cosmic ray detection capabilities, potential for the detection of dark matter annihilation or decay products, and good performance for gamma-ray astronomy. This requires to measure photons, electrons, protons and heavy ions with the same payload and large acceptance. Approved missions include CALET~\cite{ref_CALET} and ISS-CREAM~\cite{ref_CREAM} to be launched to the ISS in 2014, as well as the Chinese free-flying satellite DAMPE~\cite{ref_DAMPE}, to be launched in 2015. They aim at nuclear charge measurement by $dE/dx$ for cosmic ray physics, using plastic scintillator in CALET, Silicon Pin diodes in ISS-CREAM and plastic scintillator plus Silicon strip detectors in DAMPE. They feature calorimeters  of different technologies with a surface of order $50\times 50$ cm$^2$ and in excess of 20 radiation lengths depth. In particular, DAMPE promises to improve energy resolution for electrons and photons by a 31 $X_0$ BGO calorimeter, the thickest ever flown in space. 

\begin{figure}
\begin{center}
\includegraphics[width=0.69\textwidth]{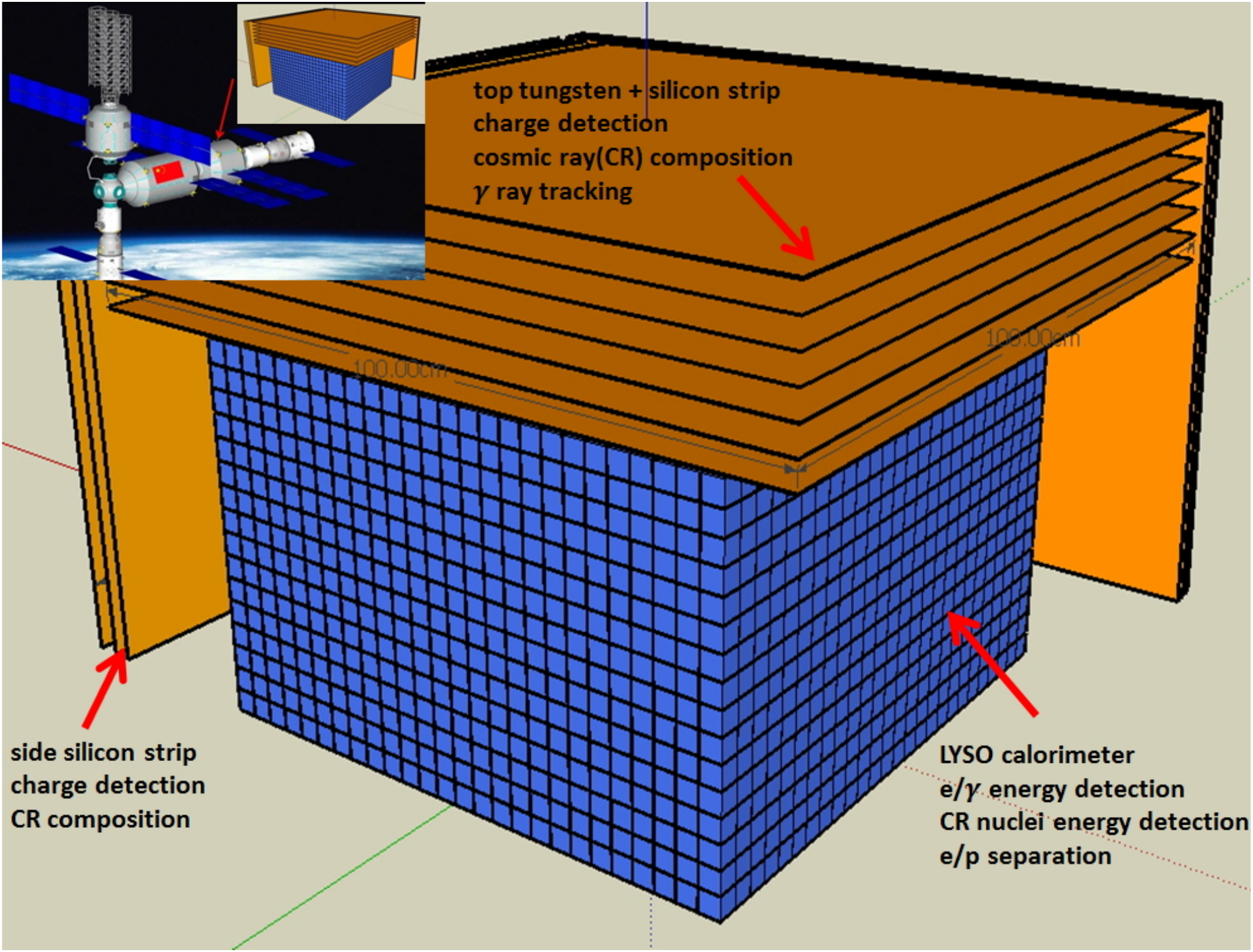}
\end{center}
\caption{Conceptual design of the layout and accommodation of HERD. The heart of the detector is a 3D calorimeter made of $3\times3\times3$ cm$^3$ Lyso cubes read out by fibers via an image intensifier. The set-up is surrounded on 5 sides by Si strip charge particle detectors, layered on top with tungsten plates for early photon conversion.}
\label{fig_HERD} 
\end{figure}

The Chinese program of scientific payloads is indeed impressively dynamic. No less than five scientific missions are in the pipeline of the Chinese space agency for the next few years, including DAMPE. Moreover, China's Space Station program has given a boost to manned missions, successfully launching, maneuvering and visiting the first prototype of the Space Laboratory Tiangong 1 since 2011. Tiangong 2 will follow in 2015, with the Chinese-European GRB polarimeter POLAR~\cite{ref_POLAR} on board. In the 2020's, a full blown Space Station will include two large modules able to host experiments. A prime candidate for an astroparticle experiment on board of the Space Station is HERD, a large 3D calorimeter surrounded by tracking devices with an acceptance of several m$^2$sr for photons, electrons and nuclei. The conceptual design of HERD is shown in Fig.~\ref{fig_HERD}. The advantage of 3D calorimetry is that it avoids a layered structure and can thus be sensitive in a field-of-view of almost $2\pi$. The intension is to reach a significant in-situ measurement of cosmic ray composition at energies approaching the "knee" at $\sim 10^{15}$ eV, where the cosmic ray spectrum changes slope for as yet unknown reasons.  

\section{Astrophysics in the keV to MeV Range}
A wide range of physics is covered by low energy space-borne photon detectors, from astrophysics to fundamental physics, thus warranting a large interest from the scientific community. Topics include the physics under extreme conditions found close to extended and compact objects like black holes or neutron stars; galactic and extragalactic cosmic rays, their origin and acceleration mechanism; the search for dark matter via self-annihilation or decay in a unique corner; and transient phenomena of all kinds. This requires excellent resolution for energy and direction, as well as polarization and timing capabilities. In this energy range, the photoelectric effect dominates at the low end, Compton scattering in the intermediate region and pair production at the high end. The region above 100 MeV is especially difficult because of the low pair production cross section. 

Several missions are proposed to cover this interesting energy range. The LOFT project~\cite{ref_LOFT} for X-ray timing in the keV range features 10 m$^2$ of Silicon Drift Detectors, inherited from ALICE at the LHC~\cite{ref_ALICE} and glass micro channel plate collimators. It has not been selected for the M3 mission of ESA's Cosmic Vision program, but will likely be a candidate for the M4 mission in the mid 2020's. In parallel, a smaller version may be accommodated earlier on the Chinese XTP mission~\cite{ref_XTP}. Among the Compton cameras proposed, the POLAR instrument~\cite{ref_POLAR} aims at measuring the polarization of photons from gamma ray bursts with unprecedented precision, using a homogeneous plastic scintillator set-up.  In the MeV to GeV range, a variety of detector concepts have been considered, aiming at reconciling the conflicting requirements of sub-degree angular resolution and large interaction probability. An interesting option seems to be a tracking calorimeter consisting of scintillating fibers without additional absorber. The enabling technology in this case is the advent of Si photomultipliers~\cite{ref_Dinu} which will allow a compact and sturdy design with high quantum efficiency. This is one of the options proposed for the PANGU~\cite{ref_PANGU} small mission concept. Simulations indicate that a point spread function of less than $1^\circ$ for $E>10$ MeV and an energy resolution of 20 to 30\% can be reached between 100 MeV and 1 GeV. 

\section{Conclusions and Outlook}
Astroparticle physics in space has entered a new era of precision measurements with AMS-02 and Fermi/LAT. Approaching TeV energies for electrons/photons and multi-TeV for ions also preduces important synergies with ground-based astroparticle experiments. Three
additional major missions, CALET, DAMPE and ISS-CREAM, will go into operation in the next two years. They aim to improve energy resolution and acceptance in the TeV range. The HERD project may well provide the next big step forward, with its large acceptance and good energy resolution up to the PeV regime. All of these offer high precision measurements of photons, electrons and ions in the same payload. Additional lever arm will come from new efforts in the keV to MeV range. Missions will observe X-ray transients, polarization and spectra relevant to better understand the origin and acceleration of cosmic rays and other energetic phenomena. New technologies for X-ray and scintillation light detection enable important progress in the field. We are thus facing an exciting program of multi-messenger, multi-wavelength astroparticle physics research in the next ten years.

\end{document}